\documentclass[12pt]{article}
\usepackage{times}
\usepackage{geometry}
\usepackage[utf8]{inputenc}
\usepackage{enumitem,amssymb}
\usepackage{ragged2e}
\newlist{thematic}{itemize}{8}
\setlist[thematic]{label=$\square$}
\usepackage{pifont}

\usepackage{graphicx}

\usepackage{natbib}

\begin{document}



{\raggedright
\huge
Astro2020 Science White Paper \linebreak

LSST Narrowband Filters \linebreak

\normalsize

\noindent \textbf{Thematic Areas:} \hspace*{60pt} $\square$ Planetary Systems \hspace*{10pt} $\square$ Star and Planet Formation \hspace*{20pt}\linebreak
$\square$ Formation and Evolution of Compact Objects \hspace*{31pt} $\boxtimes$ Cosmology and Fundamental Physics \linebreak
  $\boxtimes$  Stars and Stellar Evolution \hspace*{1pt} $\boxtimes$ Resolved Stellar Populations and their Environments \hspace*{40pt} \linebreak
  $\square$    Galaxy Evolution   \hspace*{45pt} $\square$             Multi-Messenger Astronomy and Astrophysics \hspace*{65pt} \linebreak
  
\textbf{Principal Author:}

Name: Peter Yoachim
 \linebreak						
Institution:  University of Washington
 \linebreak
Email: yoachim@uw.edu
 \linebreak
Phone:  
 \linebreak
 
\textbf{Co-authors:} (names and institutions)
  \linebreak
Melissa Graham,  University of Washington \\ 
Steven Bet, University of Washington \\ 
Milica Vu\v{c}eti\'{c}, Faculty of Mathematics, University of Belgrade\\ 
\v{Z}eljko Ivezi\'{c}, University of Washington \\ 
Martha Boyer, Space Telescope Science Institute \\ 
Bojan Arbutina, Faculty of Mathematics, University of Belgrade\\ 
Olivia Jones, UK Astronomy Technology Centre 
\linebreak

\textbf{Abstract  (optional):}
In this white paper, we present the scientific cases for adding narrowband optical filters to the Large Synoptic Survey Telescope (LSST). LSST is currently planning to observe the southern sky in 6 broadband optical filters.
Three of the four LSST science themes would benefit from adding narrowband filter observations.  We discuss the technical considerations of using narrowband filters with the LSST and lay out the scientific impact that would result on the study of AGB stars, emission line nebula (e.g., supernova remnants and planetary nebulae), photometric redshifts of galaxies, and the determination of stellar parameters.
}

\pagebreak

\section{Introduction}

The Large Synoptic Survey Telescope (LSST; for an overview see \citet{ivezic2008lsst}) is the most ambitious survey currently being constructed or planned in the visible band. 
The LSST survey power is due to its large \'etendue.
LSST will extend the faint limit of the SDSS by about 5 magnitudes and will have unique survey capabilities for faint time domain science.
The LSST design is driven by four main science themes: probing dark energy and dark matter, taking an inventory of the Solar System, exploring the transient optical sky, and mapping the Milky Way (for a detailed discussion see the LSST Science Book, LSST Science Collaboration 2009). 
LSST will be a large, wide-field ground-based system designed to obtain multiple images covering the sky that is visible from Cerro Pach\'on in Northern Chile.
The system design, with an 8.4m (6.5m effective) primary mirror \citep{Gressler2018}, a 9.6 deg$^2$ field of view, and a 3.2 Gigapixel camera \citep{Kahn2010}, will enable about 10,000 deg$^2$ of sky to be covered using pairs of 15-second exposures in two photometric bands every three nights on average, with typical 5$\sigma$ depth for point sources of $r \sim 24.5$. The system is designed to yield high image quality as well as superb astrometric and photometric accuracy. 
The LSST camera provides a 3.2 Gigapixel flat focal plane array, tiled by 189 4K$\times$4K CCD science sensors with 10 $\mu$m pixels. The current LSST plan calls for 6 broadband optical-IR filters ($ugrizy$). In this white paper, we present the scientific case for adding narrowband filters to LSST. Deep, wide field observations in narrowband filters has the potential to improve several of LSST's main science drivers.

\section{Technical Considerations}

The fast {\it f}/1.2 beam of LSST makes traditional narrowband interference filters questionable. However, the LSST beam is a large annulus, with angles of incidence between 14$^\circ$ and 23$^\circ$, leading us to believe we can achieve filter bandpasses with a FWHM of 15-20 nm, a factor of 7 smaller than the planned broadband filters.

The LSST Science Requirements Document \citep{lsstSRD} allows for about 10\% of the observing time (300 nights) to be allocated to specialized programs. If only 2 nights ($<0.1$\% of the total observing time) were allocated to a narrow-band survey, it would be possible to cover about 10,000 sq.deg. of sky in each band. Such a time allocation would match the cost of procuring the filters (of the order \$1M) to the cost of the LSST system itself (about 400,000 USD per observing night). Given that such data would also be useful for extra-galactic astronomy and cosmology, it is plausible that more than 2 observing nights could be negotiated for this program.

The LSST calibration system will include a collimated beam projector as well as a screen that can be illuminated with a tunable laser \citep{Ingraham2016}. The calibration laser covers 320-1125 nm with 1 nm increments, making it ideal for performing flat field calibration on both the standard broadband filters and potential narrowband filters. The LSST data management pipeline should have no problem with images from narrowband filters. If anything, the lower saturation level with shallower narrowband images will mean more stars in common with the {\it Gaia} catalog \citep{Gaia2018} and thus it should be easier to compute astrometric solutions.

In Table~\ref{Table:potentialFilters}, we list some potential wavelengths where narrowband filters have been used in previous studies.  Table~\ref{Table:potentialFilters} also lists the potential redshifts that bright galaxy emission line features would be detected at. 

\begin{table}
\begin{center}
\begin{tabular}{c|c|c|c|c} 
\hline
Wavelength (nm) & Local Target & z(L$\alpha$) & z([OII]) & z(H$\alpha$) \\
\hline \hline
     372 & SFRs, SNRs         & 2.1  & 0 & - \\
     436 & nebulae            & 2.6  & 0.17 & - \\
     486 & SFRs, SNRs         & 3.0  & 0.30 & - \\
     500 & nebulae            & 3.1  & 0.34 & - \\
     656 & SFRs, nebulae      & 4.4 & 0.76 & 0 \\
     672 & SNRs, superbubbles & 4.5 & 0.81 & 0.02 \\
     778 & AGB, TiO           & 5.4 & 1.1  & 0.19 \\
     812 & AGB,CN             & 5.7 & 1.2  & 0.24\\
     \hline
\end{tabular}
\caption{Potential Filters}\label{Table:potentialFilters}
\end{center}
\end{table}

\section{AGB Stars}
The progenitors of thermally pulsating asymptotic giant branch stars (TP-AGB) are red giants; their progeny are planetary nebulae and white dwarfs. 
AGB stars have a strong impact on the galactic environment; stellar winds blown during this evolutionary phase are an important component of mass return into the interstellar medium and may account for a significant fraction of interstellar dust (for a recent review, see \citet{Hofner2018}). 
These dusty winds reprocess the stellar radiation, shifting the spectral shape towards the IR. Dusty AGB stars are observable in our Galaxy and throughout the Local Group \citep{Boyer2015}, most notably the Magellanic Clouds. 
In addition to its obvious significance for the theory of stellar evolution, the study of AGB winds has important implications for the structure and evolution of galaxies \citep{Girardi2007, Maraston2009}. 
In our own Galaxy, its estimated 200,000 AGB stars are good tracers of dominant components, including the bulge \citep{Jackson2002}. 
AGB stars also have a great potential as distance indicators \citep{Rejkuba2004, Goldman2019}.

According to their photospheric chemical composition, AGB stars can be divided into oxygen-rich (O-rich) and carbon-rich (C-rich) stars; O-rich stars are associated with silicate-rich dust chemistry and C-rich stars with carbonaceous dust grains. The C-to-O stellar count ratio is an important distinguishing characteristic of stellar populations that depends on both age and metallicity; for example, it is more than two orders of magnitude higher in the Galactic center than in the Large Magellanic Cloud \citep{Nikutta2014}.
Hence, it is highly desirable for a survey to be able to classify stars into O and C classes.
At infrared wavelengths, the predominant dust type is relatively easily determined thanks to the prominent silicate dust feature at 10 $\mu$m. At optical wavelengths, it is harder to separate the two classes. The current LSST baseline design includes six broad- band filters. The separation of C and M TP-AGB stars is notoriously bad in broad-band optical fitlers.
The ability of LSST to characterize AGB stars (i.e., to separate C from O type stars) could be further enhanced by adding narrowband filters. For example, the so-called TiO (7780 \AA) and CN (8120 \AA) filters, introduced by \citet{Wing1971}, have been successfully used by a number of groups \citep[][and references therein]{Cook1986, Kerschbaum2004, Battinelli2005} for the identification and characterization of late-type stars in external Local Group galaxies.

AGB stars are not well understood and models cannot yet reproduce observations, especially at metallicity extreme. This is very important for high-redshift galaxy studies, which currently rely on poor TP-AGB models to create stellar population synthesis models, significantly biasing results since TP-AGB stars are so bright. One of the difficulties is obtaining enough C stars to be useful for calibration -- not easy with the small fields of view available with, e.g., HST. LSST will be able to get large numbers of these objects throughout the Local Group \citep{Boyer2013, Boyer2017}.

Assuming 15 nm wide filters, the faint limits would be at about apparent magnitudes 22-22.5. This is about 0.5-1 mag shallower than e.g., a study of And II by \citet{Kerschbaum2004}, but the surveyed area would be over 1,000,000 times larger! Furthermore, it is noteworthy that the deep and exceedingly accurate broad-band photometry will come for “free” and will include many epochs which can be used to reject foreground Galactic M dwarfs by variability. The same data would enable efficient calibration of the narrow-band survey.

\section{Emission Line Nebula}

Optical narrow band filters allow us to study emission-line nebulae such as HII regions, supernova remnants (SNRs) or planetary nebulae (PNe) in our Galaxy and other nearby galaxies. HII regions are one of the best-known star formation tracers in galaxies. 
Since only short-living massive stars can heat and ionize surrounding hydrogen to produce HII regions, we mainly observe such regions around OB stars and stellar associations. The ionized gas radiates intensively in H$\alpha$\ line, and therefore photometry in H$\alpha$\ filters is one of the main probes of local and global current star formation in galaxies. While supernova events happen quite suddenly and last for a relatively short time, SNRs i.e. the material ejected in an explosion continues its life through the interaction with the surrounding interstellar medium (ISM) for thousands of years. 
SNRs are responsible for enrichment of ISM with heavier elements produced in nucleosynthesis in stars and for creating turbulence in ISM from which the second generation stars have been born. Supernovae type Ia are especially important since they are widely used by astronomers as distance indicators or the so-called ”standard candles”. 
LSST and future surveys could be important in detecting runaway stars or red giant stars (or some other evolved stars from progenitor close-binary systems) which survived explosions and located in type Ia SNRs. 
Their detection can potentially give answer which scenario for type Ia explosion, single degenerate (single massive accreting white dwarf) or double degenerate (binary white dwarf merger) is prevailing.

PNe are usually detected upon their strong radiation in [OIII] doublet (4959, 5007), as well as in H$\alpha$ line (see MASH catalogue \citet{Parker2006}). LSST has the potential to produce catalogues of PNe in all nearby galaxies. Beside morphological and size criteria, as well as possible central star characteristics (which can be obtained from broad band photometry), that are used to discriminate between different types of emission nebulae, [SII] radiation is also additional possibility to separate SNRs from PNe and HII regions. Possible introduction of [OIII] 4363 narrowband filter would give us chance to determine temperatures in observed nebulae, using temperature sensitive lines ratio [OIII] (4959+5007)/4363.

\section{Photometric Redshifts}


\begin{figure}
\begin{center}
\includegraphics[scale=0.26]{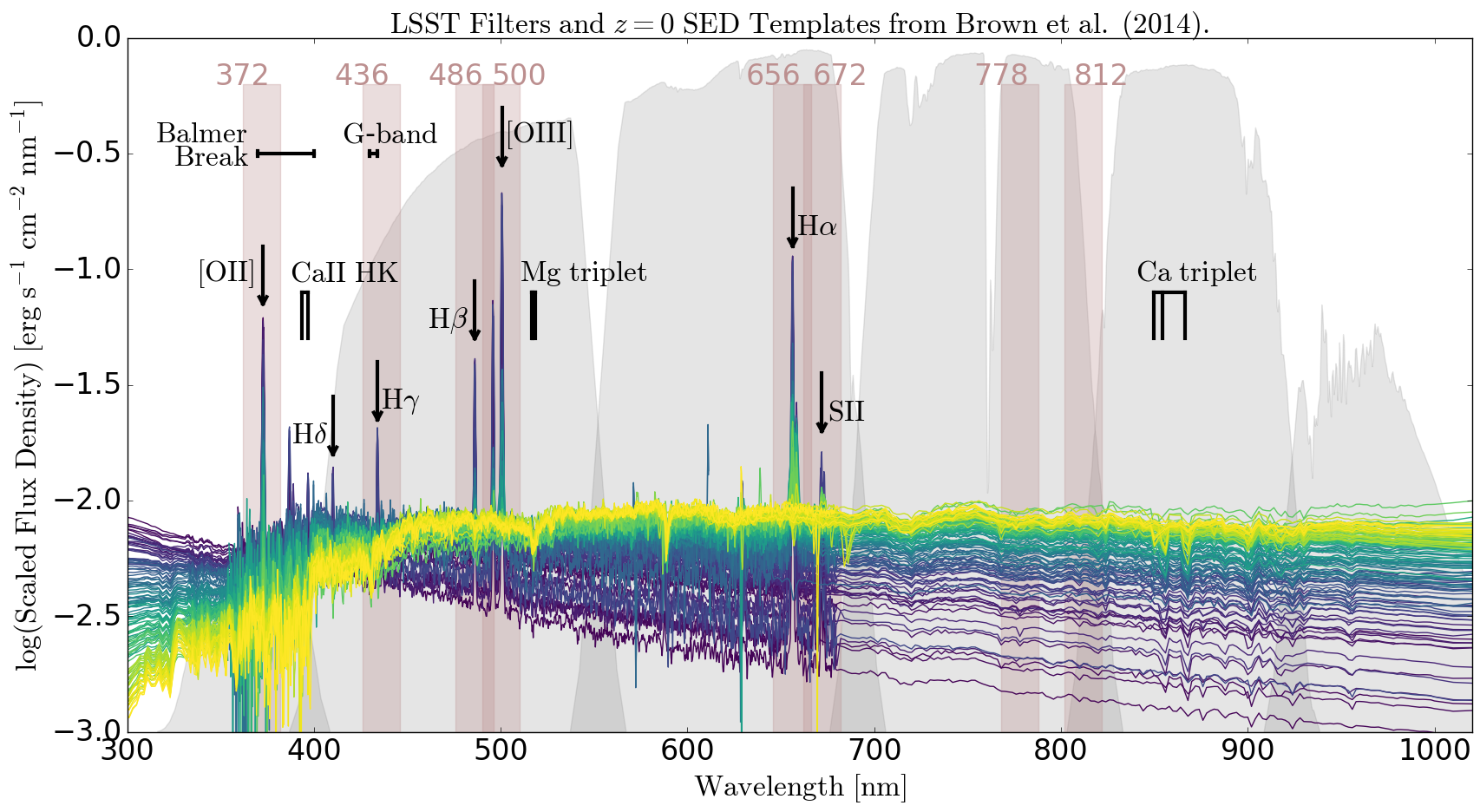}
\end{center}
\caption{The LSST broadband filters {\it ugrizy} and proposed narrowband filters (grey and rose shaded regions) are shown with $z=0$ galaxy spectral templates from \cite{2014ApJS..212...18B} (purple through yellow, ordered by $u-g$ color), for which many of the prominent spectral features have been labeled (except for the Lyman Break at $912$\ \AA, which is outside the $x$-axis range). \label{fig:galsed}}
\end{figure}

Cosmological analyses with the LSST such as weak lensing and Baryon Acoustic Oscillations (BAO) require accurate photometric redshifts for a large number of galaxies \citep{ivezic2008lsst}. Observations in narrow band filters can help to break redshift degeneracies, e.g., by identifying emission line objects at specific redshifts as features pass through the narrow filter's wavelength range. Table~\ref{Table:potentialFilters} lists the common bright galaxy emission lines and the redshifts that would be sampled by each potential narrowband filter. In Figure~\ref{fig:galsed}, we illustrate this by showing the set of $z=0$ spectral energy density (SED) templates from \cite{2014ApJS..212...18B}, some of which display emission features of hydrogen, oxygen, and sulfur, alongside the six LSST broadband filters ({\it ugrizy}) and the eight proposed narrow-band filters from Table~\ref{Table:potentialFilters}. The ${\rm [OIII]}$ emission line, for example, would enter the 656 filter at $z=0.30$ and the 672 filter at $z=0.34$, breaking a degeneracy caused by the Balmer break moving between the {\it g} and {\it r} broadband filters which degrades the photometric redshift results at $0.3 \lesssim z_{\rm phot} \lesssim 0.5$.

Narrowband filters have been shown to improve photometric redshifts. 
The COSMOS2015 catalog included two narrowband filters (at $711$ and $816$ nm central wavelengths, with $7.2$ and $12$ nm widths) and presents improved photometric redshifts for its very deep galaxy catalog (the impact of including UV and IR filters is difficult to extract from the impact of narrowband filters in this case; \citealt{2016ApJS..224...24L}).  The Javalambre-Photometric Local Universe Survey (J-PLUS), which uses 6 narrow ($<200$\  \AA\ widths) and 6 broader ($400$ to $1500$\ \AA\ widths) filters, achieves accurate photo-$z$ estimates with $\Delta z \sim 0.01$ to $0.03$ precision (for $r<20$ mag galaxies; 
\citealt{Cenarro2019}). The Physics of the Accelerating Universe Survey (PAUS), which uses 40 adjacent narrowband ($125$\ \AA\ wide) filters, achieves a photo-$z$ dispersion of $<0.002(1+z)$ (for $i_{\rm AB} <22.5$ mag galaxies; \citealt{2014MNRAS.442...92M,2019MNRAS.484.4200E}). 
However, the areas of these surveys are far smaller than the LSST wide-fast-deep main survey ($18000$\  deg$^2$): JPLUS is $36$\ deg$^2$, and PAUS will be $100$ deg$^2$ but has only released the COSMOS field ($2$\ deg$^2$) so far. Furthermore, for the LSST gold sample of galaxies with $i<25$ mag the photometry will deliver photo-$z$ estimates with dispersions of $\lesssim 0.02(1+z)$ with the broadband filters {\it alone}, so the improvements offered by narrowband imaging will start from that higher baseline. 
We leave a full simulation of the impact of narrowband filters on LSST photometric redshifts for future work.

Aside from bulk improvements on the LSST photometric redshift estimates, the narrowband filters will offer opportunities for unique cosmological analyses, such as clustering tomography in thin redshift shells across a wide area of sky to high redshift, which was previously only possible with spectroscopic catalogs (e.g., \citealt{2014MNRAS.443.3612S}). Redshift shells with significantly improved photo-$z$ estimates can also be used to minimize systematics in weak lensing studies. 

\section{Stellar Parameters}

Accurate photometric measurements of stars can be used to fit stellar surface temperatures, metallicities, and surface gravities. Such observations are crucial to understanding the star formation history of the Milky Way as well as the accretion history of stars from other systems.

\begin{figure}
\begin{center}
\includegraphics[scale=.4]{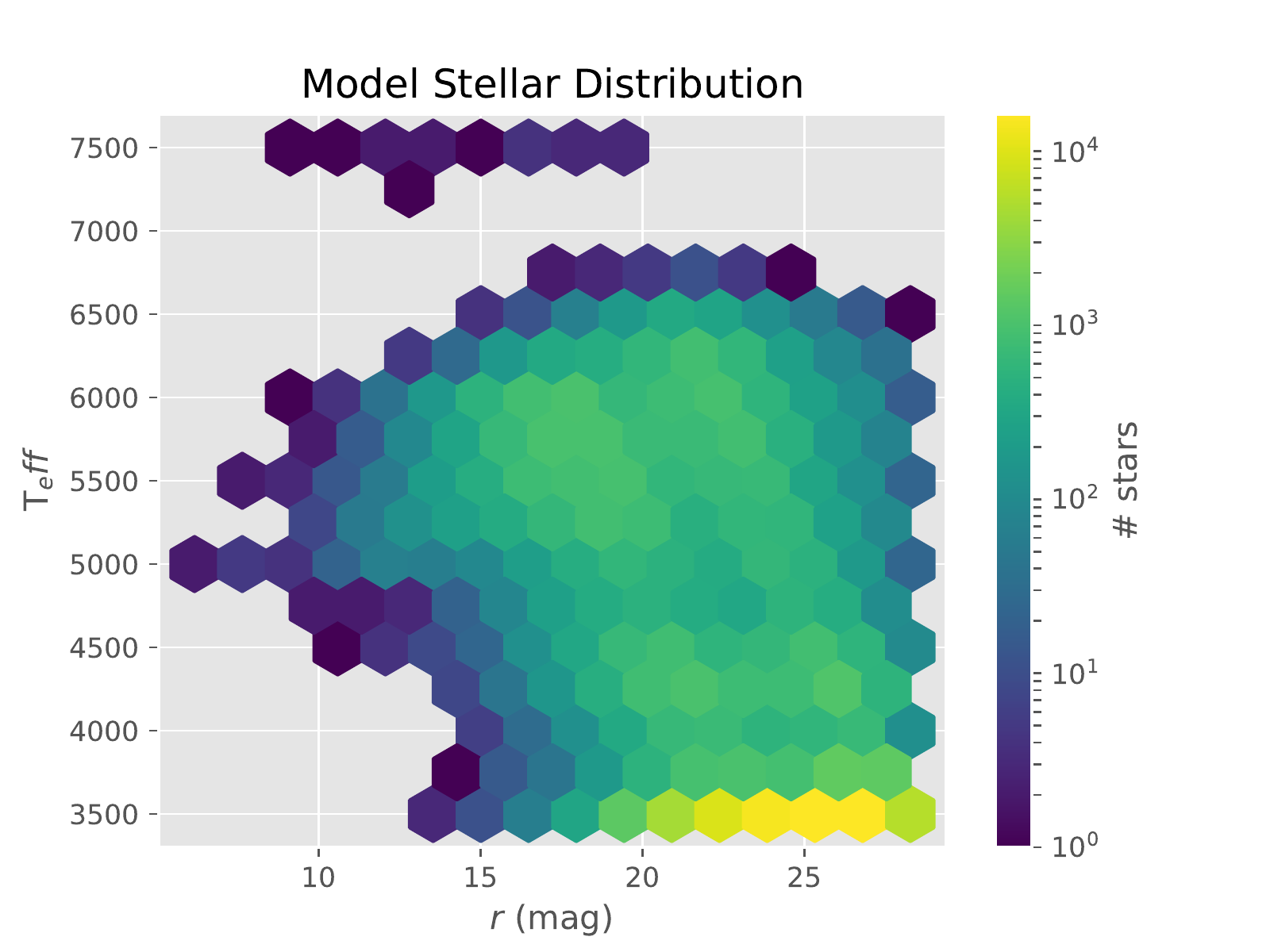} \includegraphics[scale=.4]{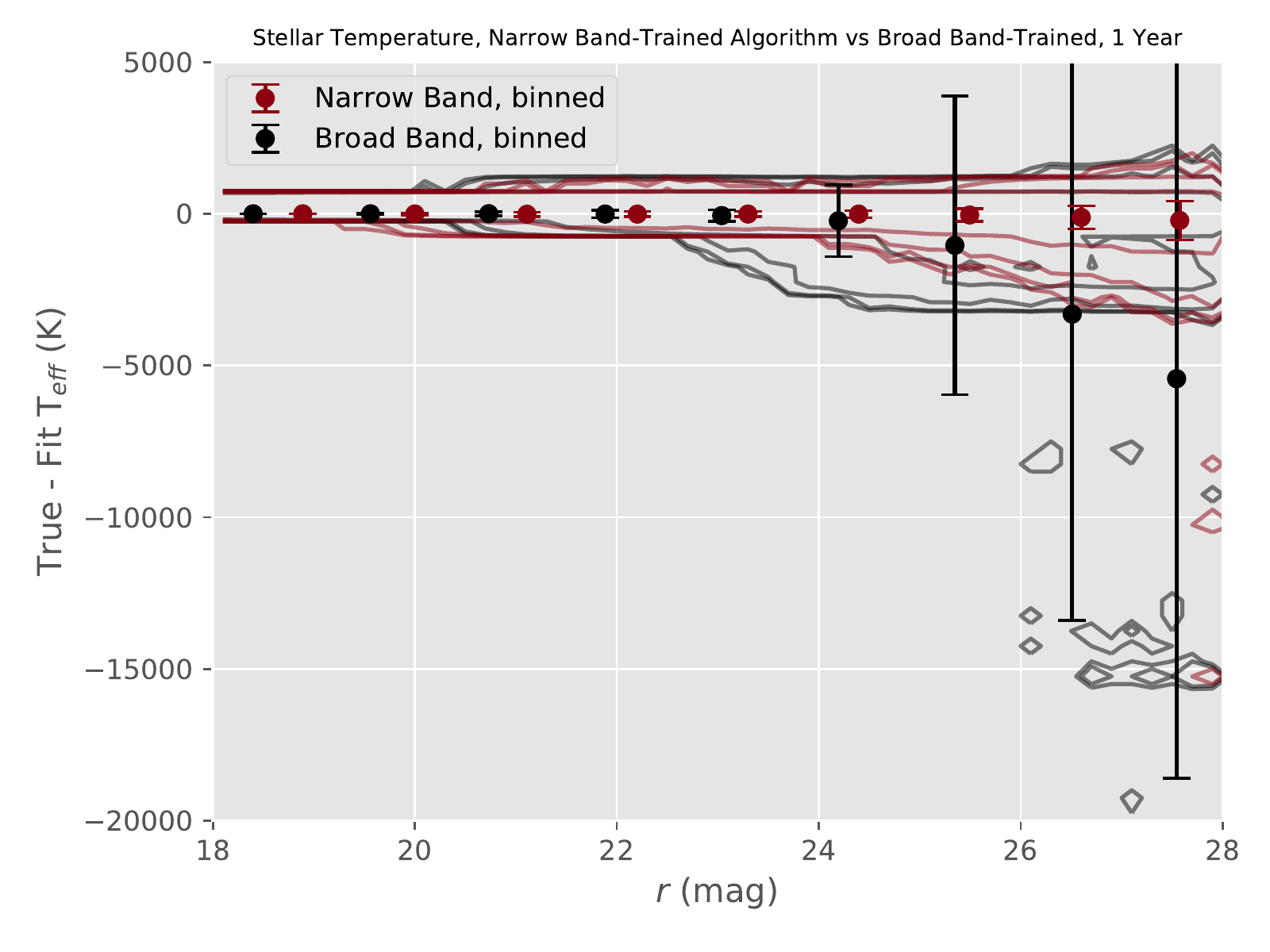}
\end{center}
\caption{ Simulation of the stars that would appear in a single high galactic latitude LSST field (left). The right panel shows how well the true stellar effective temperature can be recovered using broadband magnitudes alone (black) and then with the addition of narrowband observations in J-PLUS narrowband filters (red). Contours show the residuals for the full distribution of stars \label{fig:fitted_teff}}
\end{figure}

{\it Gaia} has measured parallaxes and proper motions for 1.3 billion stars down to $G\sim21$\ mag \citep{Gaia2018}. LSST is a natural complement to the {\it Gaia} mission and is expected to generate observations of 10 billion stars down to $r\sim27$. LSST should be able to measure accurate parallax and proper motions to $r\sim24$. Using the LSST Catalog Simulation framework \citep{Connolly2014}, we have generated a simulated LSST catalog for a small area of sky and compared the precision of fitted stellar ages and metallicities with and without additional narrowband observations.  Figure~\ref{fig:fitted_teff} shows how adding narrowband observations could improve fitted stellar parameters \citep{Bet2019}. We find that precision stellar effective temperature fits can be extended to stars 2-3 mags fainter if narrowband data is also available. For this simulation, we assumed observations taken with seven of the filters listed in Table~\ref{Table:potentialFilters}, but expect substantial gains even if only a subset of filters are used.

\clearpage

\let\oldbibliography\thebibliography
\renewcommand{\thebibliography}[1]{\oldbibliography{#1}
\setlength{\itemsep}{0pt}} 
\bibliography{LSSTbiblio}

\end{document}